# Design of an Novel Spectrum Sensing Scheme Based on Long Short-Term Memory and Experimental Validation

4 authors, including:

Nupur Choudhury
Gauhati University
2 PUBLICATIONS   0 CITATIONS

SEE PROFILE

Kandarpa Kumar Sarma
Gauhati University
450 PUBLICATIONS   1,383 CITATIONS

SEE PROFILE

Chinmoy Kalita
Gauhati University
2 PUBLICATIONS   0 CITATIONS

SEE PROFILE

Some of the authors of this publication are also working on these related projects:

Design of Complementary Split Ring Resonator based Microstrip Antenna for wireless Communication View project

Speech Processing View project

# Design of an Novel Spectrum Sensing Scheme Based on Long Short-Term Memory and Experimental Validation

[1]Nupur Choudhury, [2]Kandarpa Kumar Sarma, [3]Chinmoy Kalita, [4]Aradhana Misra

Department of Electronics and Communication Engineering,Gauhati University Guwahati, India

[1]nupur06.nc@gmail.com , [2]kandarpaks@gauhati.ac.in, [3]chinmoyk03@gmail.com,
[4]aradhana.misra@gauhati.ac.in

*Abstract—* Spectrum sensing allows cognitive radio systems to detect relevant signals in despite the presence of severe interference. Most of the existing spectrum sensing techniques use a particular signal-noise model with certain assumptions and derive certain detection performance. To deal with this uncertainty, learning based approaches are being adopted and more recently deep learning based tools have become popular. Here, we propose an approach of spectrum sensing which is based on long short term memory (LSTM) which is a critical element of deep learning networks (DLN). Use of LSTM facilitates implicit feature learning from spectrum data. The DLN is trained using several features and the performance of the proposed sensing technique is validated with the help of an empirical testbed setup using Adalm Pluto. The testbed is trained to acquire the primary signal of a real world radio broadcast taking place using FM. Experimental data show that even at low signal to noise ratio, our approach performs well in terms of detection and classification accuracies, as compared to current spectrum sensing methods.

I. INTRODUCTION

Wireless communication technologies are rapidly evolving and the growing number of wireless applications and services require addressing the spectrum shortage issue. The Federal Communications Commission (FCC), which is the United States' telecommunications authority, has shown that licenced bands are not used up to 75 to 90% of the time. The FCC Spectrum Policy Task Force published the findings of the measurement in a paper titled "FCC Report of the Spectrum Efficiency Working Group"[1]. A lot of study has been done in recent years on how to make efficient use of these spectrum bands that are either vacant or not being exploited to their full potential. With an opportunistic approach, cognitive radio (CR) is a vital technology that allows the restricted use of the inefficiently utilized frequency bands to be used more efficiently. The most important need for a CR is to detect the presence of licensed (primary) users in the spectrum, and then decide on resource allocation for unlicensed (secondary) users as a result of this detection. One of the most significant requirements for the allocation is that secondary users do not interfere with legal prime users in any way. To ensure interference-free access, secondary users must be able to consistently discover spectrum opportunities across frequency, time, and space, as well as vacate the assigned resources as soon as the primary user is active. In CR networks, whether the spectrum sensing function is executed correctly has a significant impact on communication performance and continuity.

A. Related Work

Different spectrum sensing methods have been proposed to sense limited or unused frequency bands, such as energy detection sensing [3], waveform-based sensing [4], matched filtering [5], cyclo-stationary-based sensing [7], eigenvalue-based sensing [8], and wavelet-based sensing [12]. Energy detection is a spectrum sensing approach that involves monitoring incoming signal energy and comparing it to a threshold to determine the presence or absence of the primary user. The noise power is used to determine the threshold function. Many studies have been published in order to determine the best threshold expression and increase spectrum sensing performance. If signal information such as bandwidth, operating frequency, modulation type and grade, pulse shape, and frame structure of the primary user are known, matched filtering detection approaches with shorter detection durations are favoured [9]. Cyclostationary detection uses the cyclostationarity properties of the received signals to detect primary user [7]. It detects the presence of primary users by exploiting the periodicity of the received primary signal. The principal user signals and noise power are not needed for Eigen value-based spectrum sensing [8]. The frequency bands of interest are commonly decomposed into a train of consecutive frequency sub-bands in the wavelet-based spectrum sensing approach [13]. Wavelet transform is used to detect abnormalities in these bands and determine whether the spectrum is full or empty. Although the analytical model-based techniques outlined above work well, they may not be appropriate for the real world [17].

For the needed duration, complexity, and detecting capabilities, each sensing technique offers distinct trade-offs. Machine learning (ML) and deep learning (DL) algorithms have gotten a lot of attention from industry and academia in the context of future wireless networks [14] because of their excellent learning ability using a data-driven approach and the rapid advancement in learning-based signal processing techniques [13]. Energy values and the Zhang statistic were employed as training features in a novel ANN-based hybrid sensing suggested in [16].The authors of [15] suggested an ANN-based spectrum sensing approach that uses energy and cyclostationary features to train the neural network. The authors in [22] used the same cyclostationary properties to switch to the convolutional neural networks (CNN) architecture for spectrum sensing. All of these research extract features in advance and then classify them using ANN. As a result, their performance will be heavily influenced by the advantages and disadvantages of previously derived attributes. The authors of [19] utilized spectrogram-based detection of radar emissions in the 3.5 GHz range using DNN such as CNN and recurrent neural network (RNN). It was discovered that the new detection approach outperformed the traditional one. For time-series problems, LSTM architecture, which is an upgraded form of the RNN, is preferable [21]. This is due to the fact that LSTM uses several gates in a single neuron to better synchronize previous (past time-stamps) and current information (present time-stamps) in a time series, and is thus widely utilized for temporal data. Further LSTM prevents the RNN from underperforming due to its inherent limitations. As a DL model, LSTM networks excel in learning the temporal connections in sequential data [15]. LSTM networks have been applied on wireless spectrum data in a few related studies in the literature. For example, in [22], authors presented an LSTM network-based spectrum prediction technique, while in [23], authors used an LSTM network-based modulation classification approach.

The aforementioned studies, on the other hand, have tackled the spectrum prediction problem and shown how different ML models compare in terms of accuracy. For CR networks, on the other hand, we propose an enhanced novel spectrum sensing scheme based on LSTM strategy. The detection probability has been used as a significant performance parameter. The scheme uses energy detection, likelihood ratio statistics, likelihood based goodness of fit statistics, and max-min eigenvalue statistics scheme as features to the enhanced proposed model.

*B. Contribution and Structure of the paper*

The main contribution of the paper are as follows:

- We discuss spectrum sensing as a classification problem and offer a deep learning-based spectrum sensing solution. The primary signal with features is used as the LSTM's input, and various types of signal and noise data are used to train the network.
- The proposed method is compared to other current standard spectrum sensing methods. The results reveal that our method has a higher probability of detection than other standard methods.
- The proposed model is validated using real time radio technology FM signals captured with the help of Adalm Pluto and the results found are satisfactory. It implies the model can be implemented in spectrum sensing of real time signals with satisfactory results.

The rest of the papers are as follows: In Section II, the problem formulation is described along with preliminaries of LSTM. In Section III, we present a details of the proposed model and the measurement procedure. The experimental results are presented in Section IV. In Section V finally conclusions are drawn.

## II. PRELIMINARIES

The problem of spectrum sensing can be represented as a binary hypothesis test:

$$H_1: r(n) = h\,s(n) + w(n); \quad (1)$$
$$H_0: r(n) = w(n);$$

where r(n) denotes the received signal, s(n) is the broadcast signal, w(n) W(n) is the additive white Gaussian noise (AWGN) with zero mean, and h denotes channel gain [25]. The received signal comprises just noise if no signal is present; otherwise, it also contains the sent signal. The problem described in (1) may be expressed as a classification problem with two categories from the standpoint of classification. Signal is one of the types, while noise is the other. We evaluate the spectrum sensing algorithm performance by using the detection probability and the false alarm probability:

$$P_d = Pr\{H_1|H_1\} \quad (2)$$
$$P_f = Pr\{H_1|H_0\}$$

*A. LSTM Preliminaries*

For sequential data, RNNs are a strong model. RNNs are a tight superset of feedforward ANNs, with the addition of recurrent edges that span consecutive time steps, allowing the model to have a sense of time.

RNNs are ANNs with an "underlying architecture of inter-neuronal connections that comprises at least one cycle," as defined by LSTM networks.

LSTM networks are specifically built to learn long-term dependencies and can overcome the problems that RNNs have in the past, such as vanishing and inflating gradients (Figure 1).[24-25]

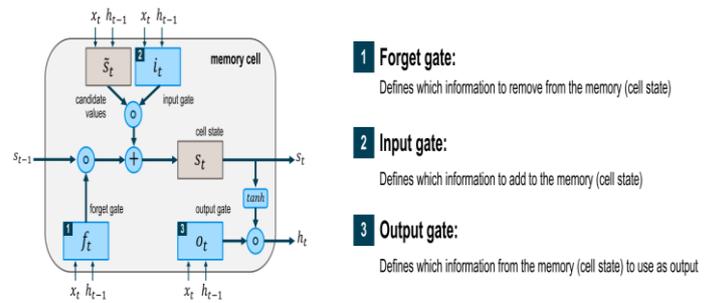

Figure 1: The architecture of LSTM memory block [24].

An input layer, one or more hidden layers, and an output layer make up an LSTM network. The number of explanatory variables (feature space) is equal to the number of neurons in the input layer. The output layer's number of neurons represents the output space, in our example two neurons signaling if a stock outperforms the cross-sectional median in t + 1. The hidden layer contains the main feature of LSTM networks (s) made up of so-called memory cells. Each memory cell contains its own set of instructions. The three gates that keep and change the state of the cell $s_t$: a forget gate ($f_t$), as well as an input gate ($i_t$) and an output gate ($o_t$). The framework a memory cell is shown in figure 2. At each timestep t, the input $x_t$ (one element of the input sequence) and the output $h_{t-1}$ of the memory cells from the previous timestep t-1 are given to each of the three gates. As a result, the gates function as filters, each serving a distinct purpose:

- The forget gate specifies which data is erased from the cell state.
- The input gate determines which data is added to the cell state.
- The output gate defines which cell state information is utilised as output.

The vectorized equations below describe the updating of memory cells in the LSTM layer at each timestep t. The following notations are used in this case:

- At timestep $t$, $x_t$ is the input vector.
- Weight matrices are $W_{f,x}$, $W_{f,h}$, $W_{ś,x}$, $W_{ś,h}$, $W_{i,x}$, $W_{i,h}$, $W_{o,x}$, and $W_{o,h}$
- Bias vectors are $b_f$, $b_ś$, $b_i$, and $b_o$.
- The activation values of the respective gates are represented by the vectors $f_t$, $i_t$, and $o_t$.
- The cell states and candidate values are represented by the vectors $s_t$ and $ś_t$
- $h_t$ is a vector that represents the LSTM layer's output.

The cell states $s_t$ and outputs $h_t$ of the LSTM layer at timestep t are determined as follows during a forward pass:

The LSTM layer selects which information should be eliminated from its prior cell states $s_{t-1}$ in the first step. As a result, the current input $x_t$, the outputs $h_{t-1}$ of the memory cells at the previous timestep (**t-1**), and the bias terms $b_f$ of the forget gates are used to compute the activation values $f_t$ of the forget gates at timestep **t**. Finally, the sigmoid function scales all activation levels to a range of 0 (totally forget) to 1 (totally recall).

$$f_t = sigmoid(W_{f,x} x_t + W_{f,h} h_{t-1} + b_f) \quad (3)$$

The LSTM layer determines which information should be added to the network's cell states($s_t$) in the second step. This technique consists of two steps: first, candidate values $s_t$ are computed that could be added to the cell states. Second, the input gates' activation values are calculated:

$$Ś_t = tanh(W_{ś,x} X_t + W_{ś,h} h_{t-1} + b_ś) \quad (4)$$

$$i_t = sigmoid((W_{i,x} X_t + W_{i,h} h_{t-1} + b_i) \quad (5)$$

The new cell states $s_t$ are determined in the third step based on the outcomes of the first two phases, with ○ designating the Hadamard (elementwise) product:

$$S_t = f_t ○ S_{t-1} + i_t ○ Ś_t . \quad (6)$$

In the final phase, the memory cells' output $h_t$ is calculated using the following two equations:

$$O_t = sigmoid(W_{o,x} X_t + W_{o,h} h_{t-1} + b_o) \quad (7)$$

$$h_t = O_t ○ tanh(S_t) \quad (8)$$

When an input sequence is processed, the LSM network is shown its features timestep by timestep. As a result, the network processes the input at each timestep t as shown in the equations. The sequence's final output is returned once the last element has been processed.

During training, the weights and bias terms are modified in the same way that they are in typical feed-forward networks to minimize the loss of the stated objective function across training samples. We pick cross-entropy as the objective function because we are working with a classification problem.

The following formula is used to determine the amount of weights and bias terms to be trained: If h represents the number of hidden units in the LSTM layer and I represents the number of input features, then the number of LSTM layer parameters that must be trained is:

$$4hi + 4h + 4h^2 = 4(hi + h + h^2) = 4(h(i + 1) + h ). \quad (9)$$

$W_{f,x}$, $W_{ś,x}$, $W_{i,x}$, and $W_{o,x}$ are the dimensions of the four weight matrices applied to the inputs at each gate, respectively. The dimensions of the four bias vectors ($b_f$, $b_ś$, $b_i$, and $b_o$) are denoted by the 4h. Finally, the dimensions of the weight matrices applied to the outputs at the previous timestep, i.e. $W_{f,x}$, $W_{ś,x}$, $W_{i,x}$, and $W_{o,x}$ are represented by the $4h^2$.

III. PROPOSED MODEL AND EXPERIMENTAL DETAILS

The most essential feature of neural networks is their ability to learn non-linear functional mappings between input and output, allowing them to adapt to the non-linear features of PU signals. The network topology is built such that the network can be trained by adopting, among others, a renowned Back Propagation[15] algorithm.

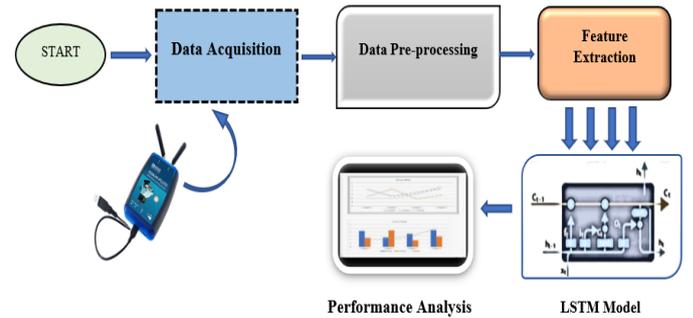

Figure 2. Proposed system model

The most essential feature of ANN/DNN is their ability to learn non-linear functional mappings between input and output, allowing them to adapt to the non-linear features of PU signals. The network topology is built such that the network can be trained by adopting, among others, a renowned Back Propagation [15] algorithm. The goal of our suggested approach is to determine whether or not the PU channel is busy or idle. We employ a supervised learning scenario in this work, where the classifier is trained given features and labels. The LSTM primarily incorporates energy detection, likelihood ratio statistics, likelihood based goodness of fit statistics, and max-min eigenvalue statistics scheme as its features. In this classifier, the labels are 0 and 1 for when the channel is idle and busy, respectively. The various statistics related to energy detection, likelihood ratio statistics, likelihood based goodness of fit statistics, and max-min eigenvalue statistics scheme are as follows:

The energy value E is shown by the discrete version of the received signal $y_i$:

$$E = \sum_{i=1}^{N}|y[n]|^2 \qquad (10)$$

The logarithm of the likelihood ratio is given by [6]:

$$L(y) = n \log \sigma_v - \log \det(\Sigma n + \sigma_v^2) - y * [(\Sigma n + \sigma_v^2 I_n)^{-1} - \sigma_v^{-2} I_n] y. \qquad (11)$$

where $I_n$ is the identity matrix.

Likelihood based Goodness of fit test (LLR-GoF):

$$Z_A = -\sum_{i=1}^{N} \left[ \frac{\log\{Fo(X(i))\}}{n-i+\frac{1}{2}} + \frac{\log\{1-Fo(X(i))\}}{i-\frac{1}{2}} \right] \qquad (12)$$

where N is the sample size and Fo is the known cumulative distribution function (CDF) of noise.

Maximum-minimum eigenvalue (MME):

$$R(N_S) = \frac{1}{N_S} \sum_{n=L}^{L-1+Ns} \hat{x}(n)\hat{x}\dagger(n) \qquad (13)$$

where † stand for Hermitian (transconjugate)

Based on the theorems [26], we have

$$\lambda_{max} \approx \frac{\sigma_\eta^2}{Ns} (\sqrt{Ns} + \sqrt{ML})^2 \qquad (14)$$

$$\lambda_{min} \approx \frac{\sigma_\eta^2}{Ns} (\sqrt{Ns} - \sqrt{ML})^2 \qquad (15)$$

where Ns is the number of collected samples. Rs is the statistical covariance matrix of the input signal, $\sigma_\eta^2$ is the variance of the noise, (M,L) is order of the matrix.

The first step is to generate simulated dataset and gather real time dataset using an empirical setup (Adalm Pluto) described in Section 3(c). The dataset is then diveded into training, validation and testing in 70:15:15 ratio ie training (70%), validation (15%) and testing (15%). Following that, we extract four features from the collected data. The scheme uses energy detection, likelihood ratio statistics (LLR), likelihood based goodness of fit statistics (LL-GoF), and max-min eignvalue statistics scheme as features to the enhanced proposed model.

The signal's energy denoted as $u_1(i)$, where i is the $i^{th}$ training sample is the first feature, followed by likelihood ratio (LLR) denoted by $u_2(i)$ the second feature, the third feature is likelihood based goodness of fit statistics denoted by $u_3(i)$ and the fourth feature is the MME denoted by $u_4(i)$. As a result, the training vector U can be written as: U= {$u_1(i)$, $u_2(i)$, $u_3(i)$, $u_4(i)$}. We employ a variety of optimization strategies in sensing with multiple hyperparameters, which may be used to increase performance in terms of detection probability.

Finally, a dense layer for the output dimension to be set according to the number of data classes is entered into the output of the last LSM cell consisting of the features correlation and dynamic characteristics of the complete sensing input sequence. We use a softmax function [14] to normalize the neural network output into [$d_{\Phi|H0}(u)$, $d_{\Phi|H1}(u)$], where $d_{\Phi|H0}(u) + d_{\Phi|H1}(u) = 1$, which stands for the opinion that the input u belongs to state H0 and H1, respectively and Φ is model parameter.

### A. Feature Extraction

To begin, data pre-processing procedures are performed on the collected data, with the data set being split into three categories: training, validation, and testing. Next to calculate the features in the training dataset N samples are extracted out wherein AWGN is added to the training dataset. For the scenario when only noise is created, feature samples of size N are also extracted in the similar way. Then the two components are sum together and each row of the dataset then serves as a detection in single instance. The training dataset contains an equal number of primary signal feature instances and AWGN feature instances to make the neural network bias-free. The labels are appended to the data collection, with label 1 denoting signal plus noise and indicating that the PU channel is busy. Similarly, label 0 denotes merely noise, meaning that the PU channel is empty.

### B. Network Training, Validation and Testing

To begin, utilizing previous knowledge of the training dataset, the features from the training dataset are computed and labelled as 0 for null hypothesis (H0 noise only) and 1 for alternative hypothesis (HA existence of primary signal) based on the sample size (N). Following that, random instances of features with their labels are retrieved and given to the LSTM model based on the batch size. The model predicts the output and improves its prediction using the back-propagation technique. The number of iterations given to the LSTM training mechanism determines how often this process occurs.

After the various LSTM model designs have been trained with the appropriate training dataset and iterations (Epoch), the best overall architecture must be chosen. Define an architecture that best suits the training set as one method to figure this out. In general, however, this does not necessarily result in a generalized model, which is why the cross-validation dataset is utilized to determine a suitable LSTM architecture. The validation data-set is used to find the model with the greatest performance on the validation data-set using various hyperparameters. The validation dataset is created from the signal data and processed in the same way that the training data is. The combined feature data is then sent to all of the LSTMs' stored models, each with its own set of hyperparameters. Different classifiers' ability to recognize correctly is identified, and then the selection stage begins. In this case, the classifier with the highest accuracy is chosen and utilized to evaluate the proposed scheme.

We move to the final phase of evaluating the sensing strategy after training and validating the suggested LSTM architecture. The testing dataset is combined with AWGN to obtain signals with necessary SNR in order to determine the detection probability $P_d$. The data is then spread into small samples (for example N=100 sample ) and features such as energy , LLR, LLR-GoF and MME statistics are retrieved. The trained network receives these properties as input. $P_d$ is calculated by dividing the number of times the neural network predicts that the channel is busy by the total number of samples for a given SNR. The probability of false alarm ($P_f$) is evaluated similarly, but instead of test data, utilizing AWGN sequences to extract features. In the output of the last LSM cell a dense layer is entered, consisting of features of the correlation and dynamic properties of the whole sensing input sequence for the output dimension to be determined according to the number of data classes. In order to normalize the neural network output, a softmax function [14] is used.

*C. Measurement Platform*

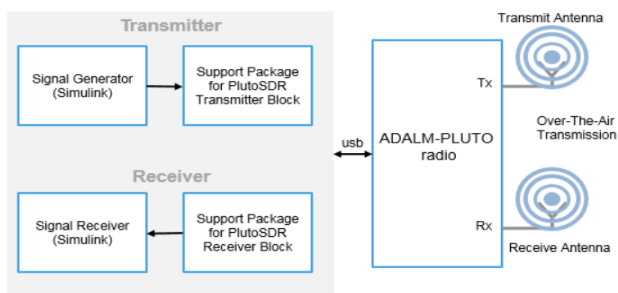

Figure 3: The following diagram shows the interaction between Simulink®, the Pluto Receiver block, and the radio hardware.

The measurement platform made use of a combination of hardware and software in this work. Analog Devices® ADALM-PLUTO (AD9361) is used as a hardware device to capture the real time primary signal data. In our study we are limited to FM broadcast signal only due to its ready availability from commercial radio stations. However the system can be configured to deal with all types of signals.

ADALM-PLUTO has one receive channel and one transmit channel that can operate in full duplex and can generate or measure RF analogue signals from 325 to 3800 MHz at up to 61.44 Mega Samples per Second (MSPS) with a 20 MHz bandwidth. It is based on the AD9363. With the default firmware, the Pluto SDR is completely self-contained, fits nicely inside a shirt pocket or backpack, and is powered entirely by USB. The FM Broadcast Demodulator Baseband System object converts the 228 kHz input sampling rate to 45.6 kHz, which is the sampling rate for the audio device. The de-emphasis lowpass filter time constant is set to 75 microseconds in the FM broadcast standard. The received mono signals are processed in this example. Stereo signals can also be processed by the demodulator. The FM Broadcast Demodulator Baseband object uses a peaking filter to perform stereo decoding by picking out the 19 kHz pilot tone from which the 38 kHz carrier is generated. The FM Broadcast Demodulator Baseband block down converts the L-R signal, centered at 38 kHz, to baseband using the resulting carrier signal. After that, the L-R and L+R signals are de-emphasized for 75 microseconds. The L and R signals are separated and converted to a 45.6 kHz.

Channels with the optimum centre frequency, decimation rate, and gain factor are chosen. The gain factor is chosen so that the received signal has the highest SNR possible, and the decimation rate guarantees that the reception bandwidth is greater than or equal to the original signal bandwidth. A total of $10^8$ samples are taken for each channel. In the initial samples, a pre-processing step is used to filter the signals and remove any abrupt peaks. The pre-processed data is then split into three data sets: training (70%), validation (15%), and testing (15%).

IV. EXPERIMENTAL RESULTS

The test findings for the proposed strategy are reported in this section. We have used Keras' library with TensorFlow backend to develop and develop models in our implementations. As a result, our trained LSTM network's topology is as follows:

- Four features and 540 timesteps in the input layer.

- LSTM layer with a dropout value of 0.1 and with h = 25 hidden neurons is created.

- This design produces roughly 3000 LSTM parameters and leads to a significant number of about 93 parameter training examples. In the event of noisy training data, a large number of observations per parameter provides for more robust estimates and decreases the risk of overfitting.

- A standard setup has two neurons in the output layer (dense layer) with a softmax activation function.

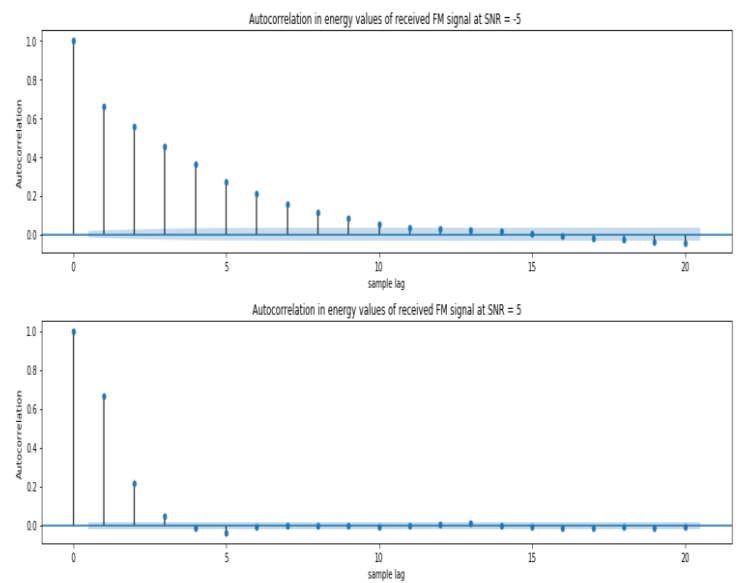

Figure 4: Autocorrelation curve of the received FM signal

For varying values of SNR, Figure 4, illustrates the autocorrelation curve of the FM signal collected using Adalm Pluto from the empirical setup. The data samples are temporally connected if the autocorrelation is not zero. This is due to the fact that the autocorrelation value does not quickly go to zero. This work uses an LSTM-based sensing framework to take advantage of the temporal correlation.

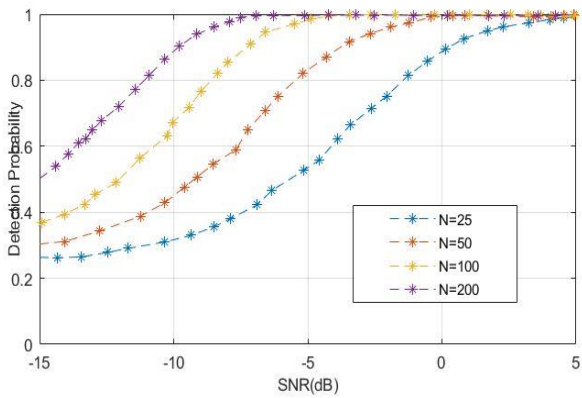

Figure 5a: Pd Vs SNR comparison for different values of N (simulation dataset)

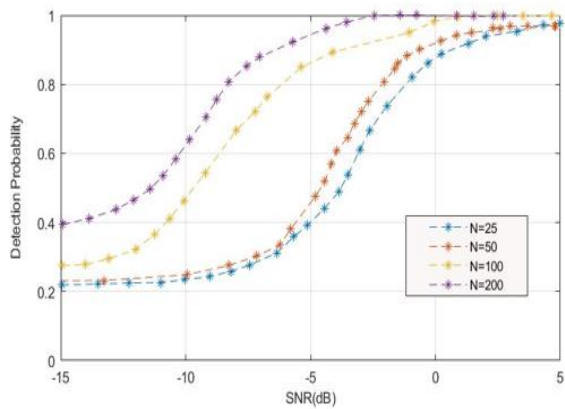

Figure 5a: Pd Vs SNR comparison for different values of N (FM signal data)

In Figures 5a and Figure 5b the impact of N on the detection probability of the proposed LSTM spectrum sensing technique for FM band data and simulation dataset is examined. It shows the $P_d$ v/s SNR curve for various values of N, where $P_d$ is detection probability. As N increases, we may note that the probability of detection also increases. In Figure 5b similar pattern is seen that implies that the model validation with real time FM signal band works in similar fashion as that of the simulated signal. Hence this model can be implemented to test other real time signal as well.

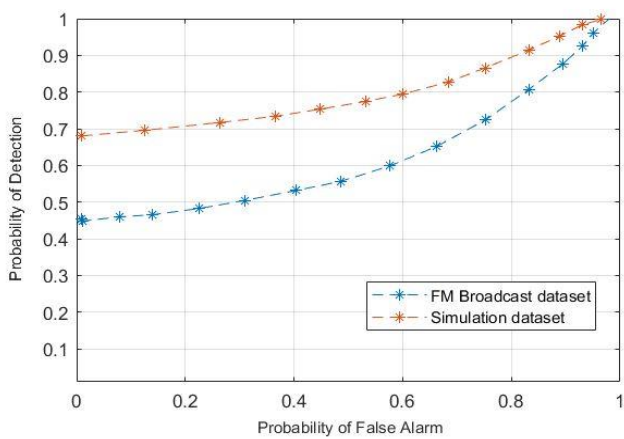

Figure 6: Pd vs Pf plot

Figure 4 depicts Pd vs. Pf operational characteristics with FM Broadcast signal dataset and simulation dataset. Pf and Pd rise in tandem with the fraction of cases in the low SNR range. The magnitudes of the PU signal are similar to noise at low SNRs. As a result, the LSTM network has trouble distinguishing between the PU signal and noise.

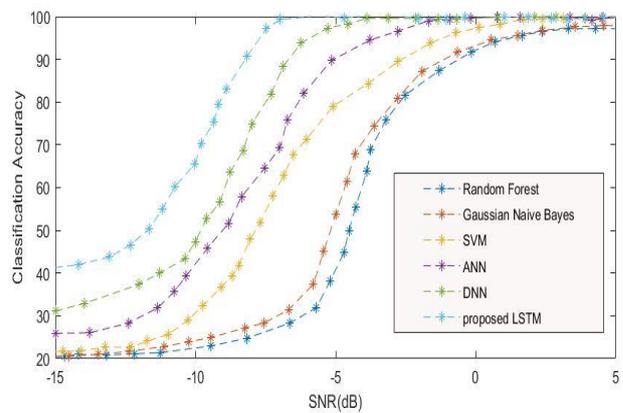

Figure 7a: Comparison of classification accuracies with other machine learning algorithms (simulation dataset)

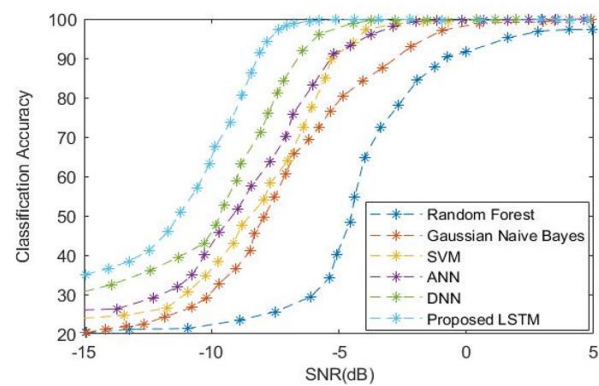

Figure 7b: Comparison of classification accuracies with other machine learning algorithms (FM signal data)

In terms of training and classification accuracy, the proposed LSTM method is compared to existing machine learning techniques such as ANN, DNN, Random forest, SVM, and Gaussian Naive Bayes. Figures 7(a),(b) compares the proposed LSTM model to various machine learning models in terms of classification accuracy. With 20 numbers of epochs, the ANN-based sensing system [17] was trained. The least number of samples required to split an internal node in the random forest classifier was two, and the tree was divided until either each leaf had one sample or all the samples in the leaves were pure. With a variance smoothing of $10^{-8}$, the Naive Bayes classifier was trained [13]. The proposed LSTM based spectrum sensing delivers a substantially enhanced classification precision compared with the different algorithms at lower SNR; but at the cost of longer training and execution periods. the sensing performance improves because LSTM cells leverage the temporal dependencies contained in the signal, which other models do not. Furthermore, the Gate structure (i.e., Update, Forget, and Output gates) regulates the information flow in LSTM, allowing it to perform effectively on temporal data. As a result, the suggested LSTM method outperform other machine learning based spectrum sensing methods. The proposed model also works approximately similar with real time FM signal band dataset that can be clearly visualize from the above graphs. Hence it implies that the proposed model is well suited to be implemented in spectrum sensing for other real time signals as well.

## V. CONCLUSION

Here we have discussed the design of a spectrum sensing scheme based on LSTM. The experimental results shows that the proposed method is suitable for application in real world situations. Further, it has better probability of correct detection and low false alarm rate compared to other existing machine learning techniques such as ANN, DNN, Random forest, SVM, and Gaussian Naive Bayes. In an extended form, the proposed model can be modified for a range of real world signal and situations.